\documentclass[twocolumn,preprintnumbers,amsmath,amssymb,superscriptaddress,nofootinbib,longbibliography,aps,pra]{revtex4-2}

\usepackage[utf8]{inputenc}
\usepackage[english]{babel}
\usepackage{graphicx}% Include figure files
\usepackage{dcolumn}% Align table columns on decimal point
\usepackage{bm}% bold math
\usepackage{amsfonts,amssymb,amsmath}
\usepackage{siunitx}
\usepackage{tablefootnote}
\usepackage{hyperref}
\hypersetup{colorlinks=true,linktoc=all,linkcolor=blue,breaklinks=true,citecolor=blue,urlcolor=blue}

\newcommand{\Qi}{$Q_\text{int}$}
\newcommand{\Qd}{$Q_\text{D}$}
\newcommand{\Qm}{$Q_\text{m}$}
\newcommand{\Qs}{$Q_\text{s}$}

\newcommand{\sr}{$\sigma_\text{released}$}
\newcommand{\er}{$\epsilon_\text{released}$}
\newcommand{\sg}{$\sigma_\text{residual}$}

\newcommand{\fm}{$f_\text{m}$}

\newcommand{\Qf}{$Q_\text{m}\times f_\text{m}$}
\newcommand{\Eref}[1]{Eq.~\ref{#1}}
\newcommand{\Fref}[1]{Figure~\ref{#1}}
\newcommand{\Tref}[1]{Table~\ref{#1}}

\newcommand{\aref}[1]{Appendix~\ref{#1}}

\DeclareSIUnit\bar{bar}

\begin{document}

\title{Thickness dependence of the mechanical properties of piezoelectric high-\Qm{} nanomechanical resonators made from aluminium nitride}

\author{Anastasiia Ciers}
\email{anastasiia.ciers@chalmers.se}
\author{Alexander Jung}
\author{Joachim Ciers}
\author{Laurentius Radit Nindito}
\author{Hannes Pfeifer}
\affiliation{Department of Microtechnology and Nanoscience (MC2),\\
Chalmers University of Technology, SE-412 96 Gothenburg, Sweden}
\author{Armin Dadgar}
\author{J{\"u}rgen Bl{\"a}sing}
\author{Andr\'e Strittmatter}
\affiliation{Institute of Physics, Otto-von-Guericke-University Magdeburg, DE-39106 Magdeburg, Germany
}
\author{Witlef Wieczorek}
\email{witlef.wieczorek@chalmers.se}
\affiliation{Department of Microtechnology and Nanoscience (MC2),\\
Chalmers University of Technology, SE-412 96 Gothenburg, Sweden}

\begin{abstract}
Nanomechanical resonators with high quality factors (\Qm{}) enable mechanics-based quantum technologies, in particular quantum sensing and quantum transduction. High-\Qm{} nanomechanical resonators in the kHz to MHz frequency range can be realized in tensile-strained thin films that allow the use of dissipation dilution techniques to drastically increase \Qm{}. In our work, we study the material properties of tensile-strained piezoelectric films made from aluminium nitride (AlN). We characterize crystalline AlN films with a thickness ranging from \SI{45}{\nano\meter} to \SI{295}{\nano\meter}, which are directly grown on Si(111) by metal-organic vapour-phase epitaxy. 
We report on the crystal quality and surface roughness, the piezoelectric response, and the residual and released stress of the AlN thin films. Importantly, we determine the intrinsic quality factor of the films at room temperature in high vacuum. We fabricate and characterize AlN nanomechanical resonators that exploit dissipation dilution to enhance the intrinsic quality factor by utilizing the tensile strain in the film. We find that AlN nanomechanical resonators below \SI{200}{\nano\meter} thickness exhibit the highest \Qf{}-product, on the order of $10^{12}$\,Hz. We discuss possible strategies to optimize the material growth that should lead to devices that reach even higher \Qf{}-products. This will pave the way for future advancements of optoelectromechanical quantum devices made from tensile-strained piezoelectric AlN.
\end{abstract}

\maketitle

\section{Introduction}
Nanoelectromechanical systems \cite{ekinci2005nanoelectromechanical} are utilized in a broad range of sensing applications, including the detection of mass \cite{chaste2012nanomechanical}, force \cite{bachtold2022mesoscopic}, or spin \cite{eichler2022ultra}. These applications are enabled by the high resonance frequency, small mass, or large mechanical quality factor of nanoelectromechanical systems. Recently, nano- and micromechanical systems have found their way into quantum technologies \cite{aspelmeyer2014cavity,barzanjeh2022optomechanics}, particularly as quantum sensors or quantum transducers. A crucial factor in these advancements was the engineering of ultra-high quality factors in these systems \cite{engelsen2024ultrahigh}. Nowadays quality factors of up to $10^{10}$ can be achieved with strained Si \cite{beccari2022strained} or SiN \cite{cupertino2024centimeter} nanomechanical resonators, with Si optomechanical crystals at ultra-low temperature \cite{ren2020two}, or with trapped SiO$_2$ nanospheres \cite{dania2024ultrahigh}. 

Dissipation dilution techniques are employed to realize chip-based mechanical resonators with ultra-high mechanical quality factors in the kHz to MHz frequency range \cite{gonzfilez1994brownian,schmid2011damping,fedorov2019generalized,engelsen2024ultrahigh}. Essentially, mechanical dissipation is reduced in tensile-strained films with a large aspect ratio by storing mechanical energy as lossless tension energy rather than as lossy bending energy \cite{fedorov2019generalized}. The increase in quality factor is quantified by the dilution factor, which dilutes the material-dependent intrinsic quality factor.

Dissipation dilution techniques have been applied to amorphous as well as crystalline materials. The latter could offer larger intrinsic quality factors due to a lower defect density. A major advantage of crystalline materials is the possibility for in-built functionality, which can be used, for example, for read-out or actuation of mechanical motion. In particular, piezoelectric films offer the possibility to directly couple mechanical with electrical degrees of freedom, and when also interfacing to light, hybrid optoelectromechanical quantum devices can be realized \cite{midolo2018nano}. Tensile-strained piezoelectric materials that have been used so far to realize high-\Qm{} nanomechanical resonators were InGaP \cite{cole2014tensile,buckle2018stress,manjeshwar2023high}, SiC \cite{romero2020engineering,sementilli2024ultralow}, and AlN \cite{ciers2024nanomechanical}. AlN is widely used in classical technologies, for example, in telecommunication due to its high acoustic velocity, piezoelectricity \cite{sinhaPiezoelectricAluminumNitride2009}, low dielectric loss, and mechanical strength \cite{tsubouchi1981aln}. Applications include multi-frequency bandpass filters, parametric feedback oscillators, and gas sensors \cite{villanuevaNanoscaleParametricFeedback2011,piazzaSingleChipMultipleFrequencyALN2007,ivaldi201150}. In quantum technologies, non-strained AlN resonators operating at GHz frequencies have for example been coupled to superconducting qubits \cite{o2010quantum}. We have recently demonstrated that a highly-tensile strained \SI{295}{\nano\meter}-thick AlN film can be used to realize nanomechancial resonators with a \Qm{} exceeding $10^{7}$ \cite{ciers2024nanomechanical}, opening the possibility to realize AlN-based high-\Qm{} optoelectromechanical quantum devices.

\begin{figure*}[t!hbp]
    \centering
    \includegraphics[width=\textwidth]{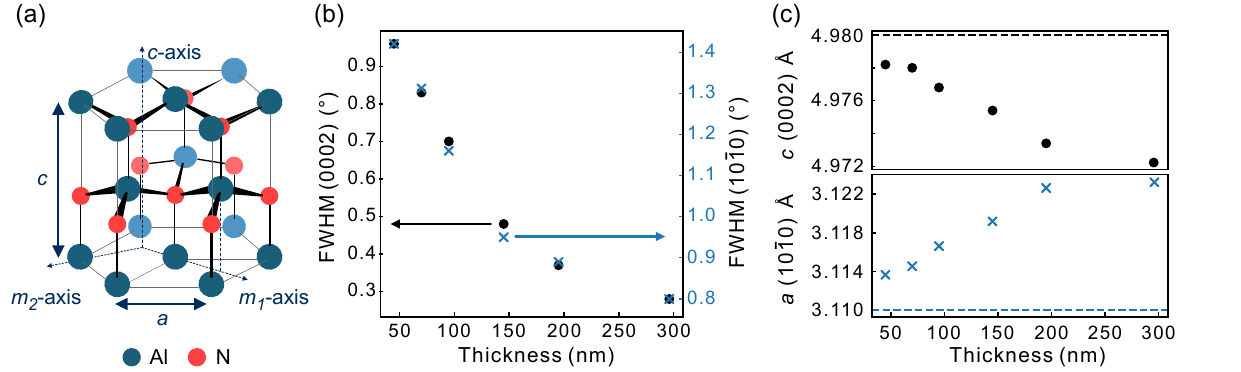}
    \caption{Crystal structure of AlN thin films. (a) Illustration of the wurtzite AlN crystal, where the $m_1$ and $m_2$ axes point along two mirror planes of the crystal. (b) HRXRD measurements: FWHM of (0002) and (10$\bar{1}$0) $\omega$-scans as a function of AlN thickness. (c) Lattice constant: (0002) corresponds to the $c$-axis (top panel) and (10$\bar{1}$0) corresponds to the $m$-axis (bottom panel). Dashed lines are the single-crystal values.}
    \label{fig:material}
\end{figure*}

The present work systematically studies the material properties of tensile-strained AlN thin films for realizing high-\Qm{} nanomechanical resonators. We study AlN thin films with a thickness between \SI{45}{\nano\meter} to \SI{295}{\nano\meter} that are epitaxially grown on (111)-oriented Si wafers by metal-organic vapour-phase epitaxy (MOVPE). Our study is motivated by the fact that the dilution factor is expected to increase for thinner films. This would result in an increase of \Qm{} \cite{fedorov2019generalized} provided that the film's residual strain and intrinsic quality factor is not reduced. We characterize the crystal structure quality of the AlN films, their surface roughness, and evaluate the residual stress of the thin films. We then fabricate AlN cantilevers to determine the Young's modulus of the AlN films. We find that the released cantilevers exhibit bending, indicating inhomogeneous strain within the film. This motivates us to introduce a simple bilayer model, which we further use in finite element method (FEM) simulations of various nanomechanical resonator geometries. Importantly, we determine the intrinsic quality factor and released stress of each AlN film. Finally, we fabricate a high-\Qm{} triangline-shaped nanomechanical resonator in the \SI{90}{\nano\meter}-thick AlN film, the thinnest AlN film that withstands the release step of the fabrication process. This triangline resonator exhibits a \Qf-product for its fundamental mode of $3.7\times 10^{12}$\,Hz, a factor of two larger than what would be expected from a similar triangline in \SI{295}{\nano\meter}-thick AlN.

%%%%%%%%%%%%%%%%%%%%%%%%%%%%%%%%%%%%%%%%%%%%%%
\section{Material characterization}
%%%%%%%%%%%%%%%%%%%%%%%%%%%%%%%%%%%%%%%%%%%%%%
\subsection{Crystal structure}
%%%%%%%%%%%%%%%%%%%%%%%%%%%%%%%%%%%%%%%%%%%%%%
\label{subsec:structure} 

Silicon substrates used for the epitaxial growth of III-nitrides (InN, GaN, AlN and their alloys) offer significant technological advantages over other growth substrates such as scalability and the ability to integrate III-nitride devices with silicon microelectronics on a single wafer.
We grow our AlN film by metal-organic vapour-phase epitaxy (MOVPE) on 2-inch (111)-oriented \SI{500}{\micro\meter}-thick Si wafers (for details see \aref{app:growth}). The resulting wurtzite AlN (see \Fref{fig:material}(a)) has a polar axis along the [0001] direction, i.e., the $c$-axis \cite{ciers2024nanomechanical}.

The bond lengths of AlN and Si differ by 16.7\%, which leads to a lattice mismatch and a high defect density in the lower \SI{70}{\nano\meter} region of the AlN film. 
III-nitrides grown on a Si substrate host all major defect types: point defects, line defects, plane defects and inversion domains \cite{li2010iii}.

The crystallinity and dislocation density of the AlN film were analyzed by high resolution X-Ray diffraction (HRXRD), details see \aref{app:char}.
To this end the detector angle is fixed and the incident X-ray beam angle is scanned ($\omega$-scan) around the (0002) or the (10$\bar{1}$0) plane. 
A narrower full width at half maximum (FWHM) of the observed diffraction peak corresponds to less disorder of the crystal plane. 
We evaluate the $c$-axis disorder using the $\omega$-scan of the AlN (0002) peak, while the twist distribution around the $c$-axis is evaluated via the (10$\bar{1}$0) peak, as shown in \Fref{fig:material}(b). 

The AlN quality clearly improves with increasing thickness \cite{mastro2006mocvd}, as the FWHM of the corresponding peaks decreases with an increase in film thickness. 
\Fref{fig:material}(c) shows the lattice constants $a$ and $c$, which are calculated from the HRXRD $2\Theta/\theta$ data using the Bragg law and interplanar spacing formula \cite{schuster1999determination}.
The observed lattice constant evolution with AlN thickness can be explained as follows.
Initially, there is a high density of small low-strain AlN crystallites. With increasing film thickness these crystallites coalesce and the film quality improves due to a reduction in grain boundaries and dislocations.
This is accompanied by an increased stress in the film \cite{nix1999crystallite}. 
Thus, the lattice constants are deviating stronger from the single-crystal values for thicker AlN films.

As the AlN film on the Si substrate is grown at high temperatures (above \SI{900}{\degree}C, see \aref{app:growth}) after cooldown a large tensile stress is generated due to the thermal expansion mismatch between Si and AlN \cite{marchand1999structural}.
The resulting biaxial tension could be partially relaxed if the crystal structure would posses a plane along which the dislocations can move. For wurtzite III-nitride materials the primary slip plane is (0001). However, breaking the covalent Al-N bonds requires significant energy \cite{jin2004growth} making AlN brittle and resistant to such plastic deformation. 
Ref.~\cite{mastro2006mocvd} showed that a critical AlN film thickness of around \SI{300}{\nano\meter} exists beyond which the AlN film starts cracking at misfit dislocations. 
We observe that our \SI{295}{\nano\meter}-thick AlN film exhibits cracks at the edge of the wafer, i.e., this film thickness appears to be the thickest to be usable for realizing AlN nanomechanical resonators. 

\begin{figure*}[t!hbp]
    \centering
    \includegraphics[width=\textwidth]{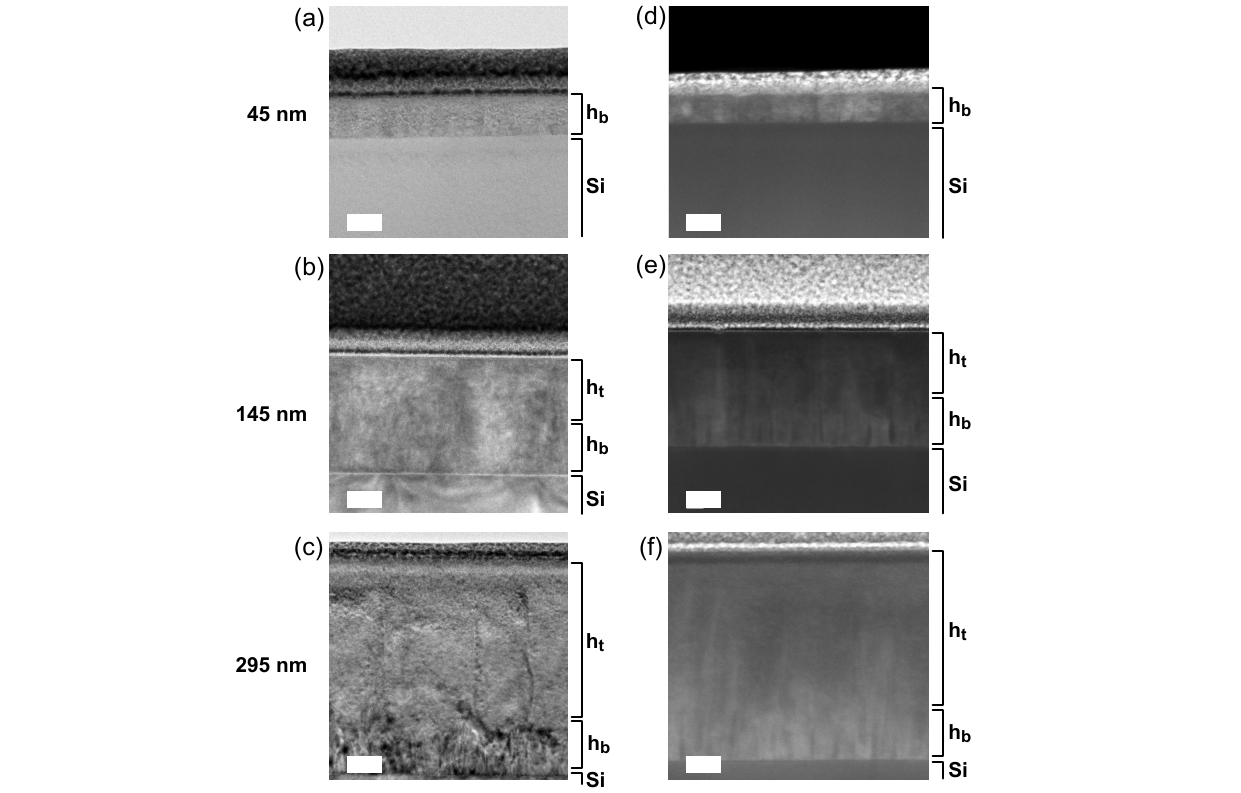}
    \caption{STEM and TEM imaging of AlN thin films. (a-c) STEM images (labeled by thickness determined through ellipsometry at the center of the wafer), (d-f) TEM images of AlN films on a Si (111) substrate. The scale bar is \SI{50}{\nano\meter}.}
    \label{fig:TEM}
\end{figure*}

%%%%%%%%%%%%%%%%%%%%%%%%%%%%%%%%%%%%%%%%%%%%%
\subsection{Film morphology}
%%%%%%%%%%%%%%%%%%%%%%%%%%%%%%%%%%%%%%%%%%%%%

To determine the thickness and the lateral variation of the AlN films, we perform spatially-resolved ellipsometry measurements across the wafer.
For the nominally \SI{295}{\nano\meter}-thick AlN film, we observe that the film thickness decreases to \SI{280}{\nano\meter} at the edge of the 2-inch wafer. For all following analysis, we therefore use chips taken from the center of the wafer and refer to the AlN film thickness at the wafer's center.

We performed transmission electron microscopy (TEM) and scanning transmission electron microscopy (STEM) imaging to evaluate the microscopic quality of the MOVPE-grown AlN film. All of the III-nitrides have smaller interatomic spacing within their basal planes compared to (111) silicon as well as higher thermal expansion coefficients. These two properties result in an interface layer with a high density of misfit dislocations \cite{radtke2012structure}. 
In the TEM images shown in \Fref{fig:TEM} such a defect-rich layer \cite{mante2018proposition} at the interface of the AlN film and the Si substrate is observed. From the change in morphology of the AlN films that are thicker than \SI{100}{\nano\meter} (see \Fref{fig:TEM}(b,c,e,f)), we estimate that this defect-rich layer is about \SI{70}{\nano\meter}.
This motivates us to introduce a simple bilayer model, which we will use to model the properties of nanomechanical resonators. This model accounts for the crystal structure difference between two layers, but neglects the gradual transition of the AlN film's crystal structure from polycrystalline to single-crystal-like. The first region close to the substrate, denoted as $h_\text{b}$, consists of many dislocations and misoriented domains and is about \SI{70}{\nano\meter} thick. Later, we fix the thickness of the defect-layer by fitting the bilayer model to measurements of fabricated cantilevers. The second region, denoted as $h_\text{t}$, is an AlN layer of high-crystal quality. Note that the emergence of a defect-rich layer in the epitaxial growth of AlN on Si (111) substrate is similar to epitaxially-grown SiC on a Si substrate \cite{romero2020engineering}.

%%%%%%%%%%%%%%%%%%%%%%%%%%%%%%%%%%%%%%%%%%%%%
\subsection{Surface roughness}
%%%%%%%%%%%%%%%%%%%%%%%%%%%%%%%%%%%%%%%%%%%%%

\begin{figure*}[t!hbp]
    \centering
    \includegraphics[width=\textwidth]{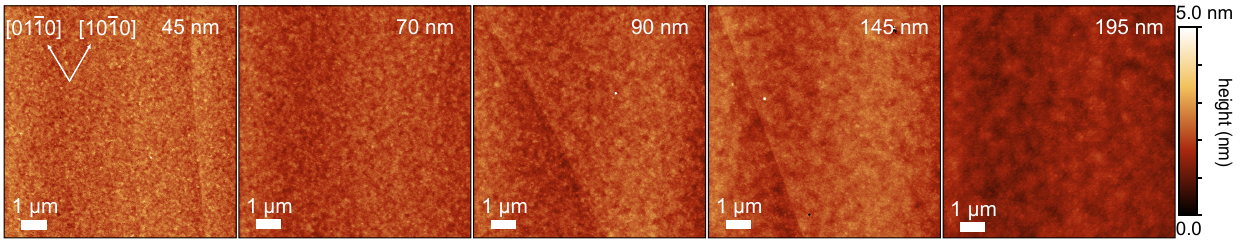}
    \caption{$10\times \SI{10}{\micro\meter}^2$ AFM maps of AlN films. The crystal orientation (same for all scans) is indicated in the first map.}
    \label{fig:AFM}
\end{figure*}

The surface roughness of the AlN layer is measured by atomic force microscopy (AFM) in tapping mode for a scan size of $10 \times \SI{10}{\micro\meter^2}$ (see \Fref{fig:AFM}). For each AlN thickness the measurements are performed at several locations of the wafer and the average root-mean-square (RMS) surface roughness is determined (see \Tref{tab:AFM_h}).
A smooth surface is essential to obtain low optical and mechanical loss.
When the films have RMS surface roughness well below \SI{1}{\nano\meter} then the optical scattering loss is small and the film is excellent for photonic applications. A low surface roughness is also an indicator of a surface with few dangling bonds and defects and, as a result, a larger intrinsic mechanical quality factor \cite{mohanty2002intrinsic}. We compare the surface roughness of unprocessed and processed AlN films of different thickness in \Tref{tab:AFM_h}. We observe that the surface roughness of all AlN films is below \SI{1}{\nano\meter}, even after fabrication of devices. However, the RMS is slightly higher in the latter case, which could be due to the etching steps in the fabrication process.

\begin{table*}[h!tbp]
    \centering
    \begin{tabular}{c c c c c c}
        \hline
         $h$\,(nm) & 45 & 70 & 90 & 145 & 195 \\
         \hline
         RMS as-grown\,(pm) & 355 $\pm$ 32 & 270 $\pm$ 6 & 402 $\pm$ 83 & 348 $\pm$ 30 & 244 $\pm$ 37 \\
         RMS processed\,(pm) & \textemdash & \textemdash & 521 $\pm$ 6 & 384 $\pm$ 6 &  408 $\pm$ 12 \\
        \hline
    \end{tabular}
    \caption{Surface roughness of AlN thin films. RMS roughness of the AlN films over a $10 \times \SI{10}{\micro\meter^2}$ scan window before and after fabrication of devices.} 
    \label{tab:AFM_h}
\end{table*}

%%%%%%%%%%%%%%%%%%%%%%%%%%%%%%%%%%%%%%%%%%%%%
\subsection{Piezoelectricity}
%%%%%%%%%%%%%%%%%%%%%%%%%%%%%%%%%%%%%%%%%%%%%

\begin{figure*}[t!hbp]
    \centering
    \includegraphics[width=\textwidth]{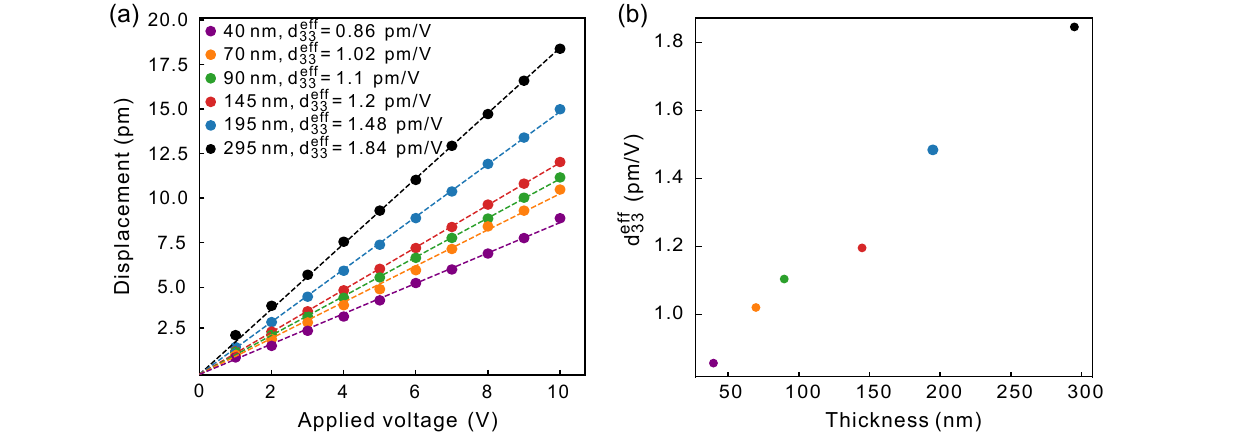}
    \caption{Piezoresponse force microscopy. (a) Piezoelectrically-induced displacement vs.~applied voltage (dashed line is a linear fit). (b) Mean values and uncertainty of the effective piezoelectric coefficient vs.~thickness of the AlN film. Note that the uncertainty is in most cases smaller than the point size.}
    \label{fig:PFM}
\end{figure*}

AlN has three piezoelectric coefficients $d_{31}$, $d_{15}$, and $d_{33}$. Among them, the piezoelectric coefficient $d_{33}$ is commonly used to quantify the piezoelectricity of AlN \cite{ciers2024nanomechanical}. We use piezoresponse force microscopy (PFM) to determine the effective piezoelectric coefficient $d_{33}^\text{eff}$ (see Ref.~\cite{ciers2024nanomechanical} for details of the measurement procedure performed with a commercial Bruker Dimension ICON AFM). PFM is based on the inverse piezoelectric effect. An AC modulation signal is applied to the surface of the AlN film through the PFM probe in contact mode. The deformation of the film induced by the electric field is detected by the PFM probe, see \Fref{fig:PFM}(a). The measurement results are effective piezoelectric values which are smaller than $d_{33}$ due to the clamping constraint of the film through the substrate \cite{kim2005study} and, potentially, the stress in the film \cite{berfield2007residual}. \Fref{fig:PFM}(b) shows that the effective piezoelectric coefficient increases with the thickness of AlN. This behavior can be explained by the improved crystal quality of thicker AlN films, which, thus, have larger polarized domains.

\begin{figure*}[ht!]
    \centering
    \includegraphics[width=\textwidth]{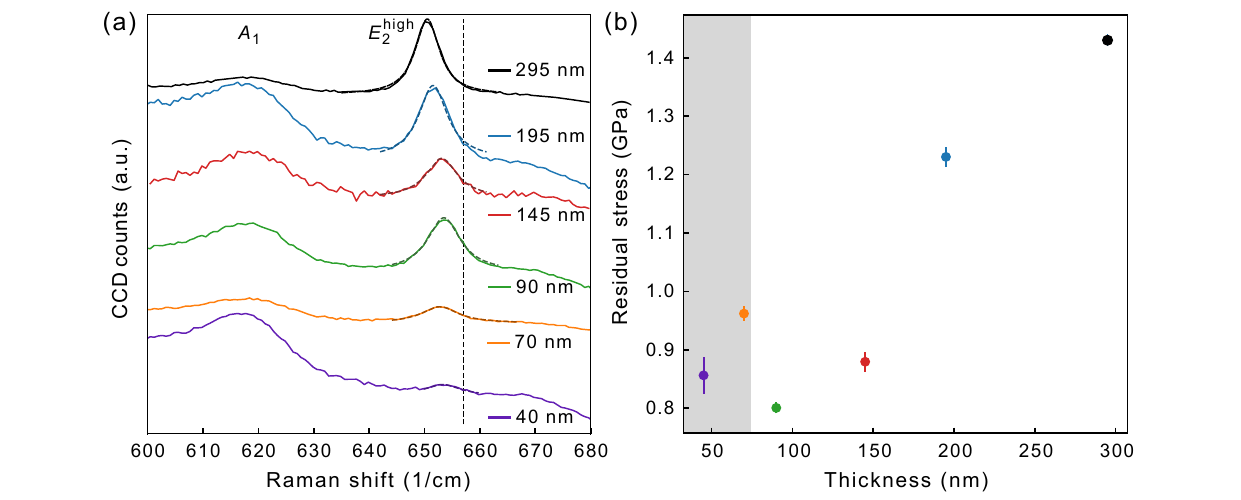}
    \caption{Raman measurements to determine residual stress. (a) Raman spectra of AlN thin films. The position of the $E_2^{\text{high}}$ phonon mode is fitted. Dashed black line at 657\,cm$^{-1}$ is the single-crystal value. (b) Evaluated \sg{} of the AlN film vs.~thickness. The grey area marks the thickness of the defect-rich layer region.}
    \label{fig:Raman_sg}
\end{figure*}

%%%%%%%%%%%%%%%%%%%%%%%%%%%%%%%%%%%%%%%%%%%%%
\subsection{Residual stress}
%%%%%%%%%%%%%%%%%%%%%%%%%%%%%%%%%%%%%%%%%%%%%

% and is polarized in the $z$-direction ($c$-axis)
%The broad lower energy mode is related to the $A_1$ branch \cite{perlin1993raman,kazan2006temperature}. %(TO)

A method to assess the residual stress of the film is to perform Raman measurements \cite{ciers2024nanomechanical}. \Fref{fig:Raman_sg}(a) shows Raman spectra for the AlN films.
Between 600 to 680\,1/cm we observe the two Raman-active phonon modes $A_1$ and $E_2^\text{high}$ \cite{perlin1993raman,kazan2006temperature}, which both shift due to residual stress. We choose to use the clear and strong signal given by mode $E_2^\text{high}$ to infer the stress in the AlN layer (see also Ref.~\cite{placidi2009highly,callsen2014phonon}) as mode $A_1$ shows a broader peak and the analysis is more complex due to its polar nature. 
The shift of the $E_2^\text{high}$ phonon mode in AlN, $\Delta\omega$, is related to the residual stress via \cite{callsen2014phonon}
\begin{equation}
    \Delta\omega = 2\alpha_0 \sigma_\text{residual}.
\end{equation}
The conversion of $\Delta\omega$ to stress in the film depends on the deformation potential constant, $\alpha_{0}$, which varies for each AlN thickness due to the presence of the defect-rich layer.
To get an estimate of the residual stress in the AlN thin films, we use the bulk single-crystal value of AlN ($\alpha_{0} = \SI{2.2115}{\centi \meter^{-1} / \giga \pascal}$ \cite{dai2016properties}). \Fref{fig:Raman_sg}(a) shows the Raman spectra and \Fref{fig:Raman_sg}(b) the evaluated residual stress of the AlN thin films. We observe that \sg{} monotonously increases with film thickness for films thicker than \SI{90}{\nano\meter} with a maximal value of \SI{1.4}{\giga\pascal} for the \SI{295}{\nano\meter}-thick film. Thin films below \SI{90}{\nano\meter} show a different behavior, which we attribute to the \SI{70}{\nano\meter}-thick defect-rich layer, which is expected to have different elastic properties than the single-crystal case.

\begin{figure*}[t!hbp]
    \centering
    \includegraphics[width=\textwidth]{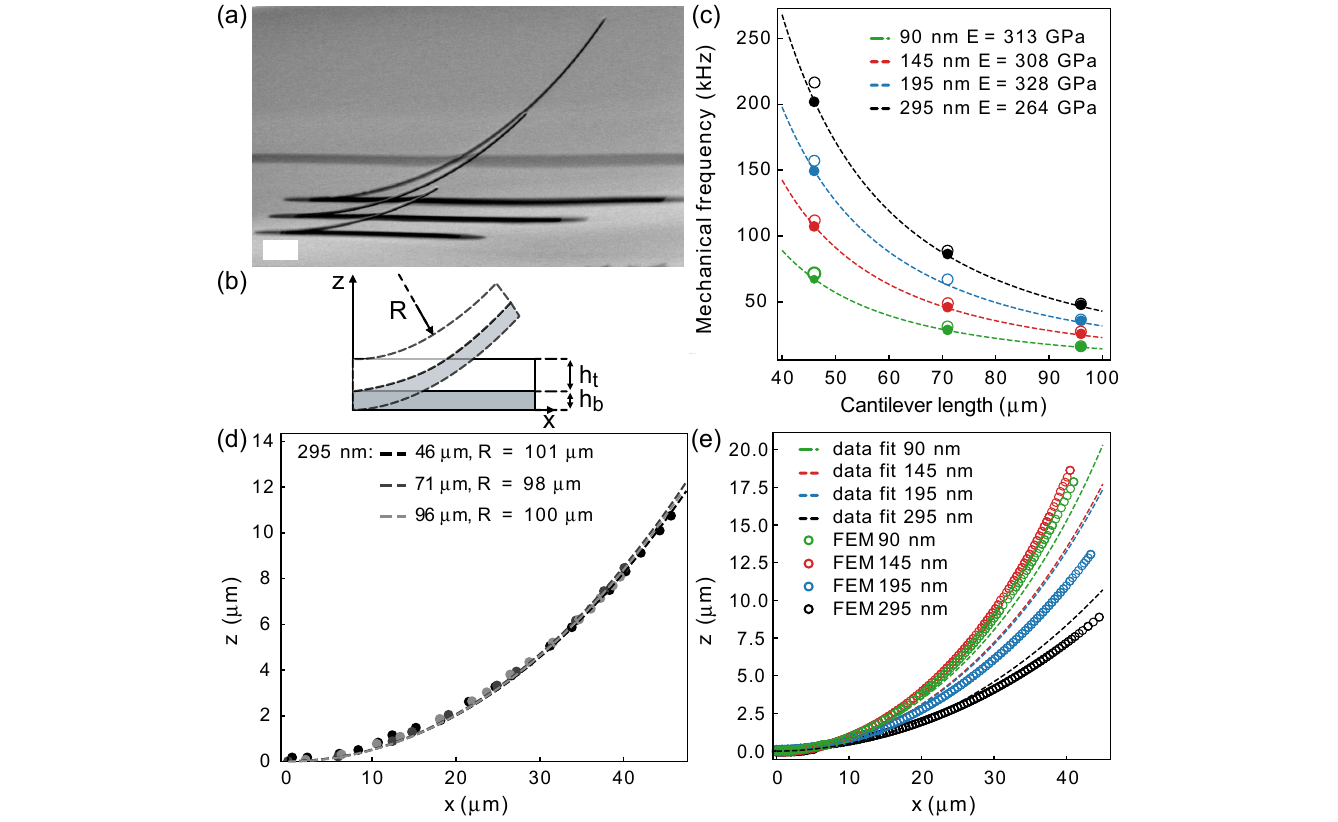}
    \caption{AlN cantilevers. (a) Tilted SEM image of \SI{46}{\micro\meter}-, \SI{71}{\micro\meter}-, and \SI{96}{\micro\meter}-long cantilevers fabricated from \SI{295}{\nano\meter}-thick AlN. The scale bar is \SI{10}{\micro\meter}. (b) Sketch of the bilayer model.
    (c) Fundamental mechanical mode frequency for cantilevers of different length and thickness. The dashed line is a fit to \Eref{eq:f_c}, dots are the measured data, and empty circles are the FEM results using the bilayer model.
    (d) Confocal microscope data (dots) and fit (dashed lines) for \SI{295}{\nano\meter}-thick AlN cantilevers of different length. 
    (e) Bending of \SI{46}{\micro\meter}-long cantilevers for \SI{90}{\nano\meter}-, \SI{145}{\nano\meter}-, \SI{195}{\nano\meter}- and \SI{295}{\nano\meter}-thick AlN films.
    The dashed lines are fits to confocal microscope data with the values from \Tref{tab:h_cant} and open dots are FEM results using the bilayer model.
    }
    \label{fig:SEM_cant}
\end{figure*}

%%%%%%%%%%%%%%%%%%%%%%%%%%%%%%%%%%%%%%%%%%%%%
\section{Cantilevers}
%%%%%%%%%%%%%%%%%%%%%%%%%%%%%%%%%%%%%%%%%%%%%
We fabricated cantilevers of varying lengths $L$ (\SI{46}{\micro\meter}, \SI{71}{\micro\meter}, and \SI{96}{\micro\meter}) to determine Young's modulus of the AlN thin films (details of the fabrication are found in \aref{app:fab}). \Fref{fig:SEM_cant}(a) shows that the cantilevers bend upwards upon release. This bending can be explained by an inhomogenous strain in the AlN film, which originates from the movement of edge dislocations and their annihilation as the AlN layer becomes thicker. To model the observed bending, we use the bilayer model. We assume that the two layers of the bilayer model exhibit a difference in strain, $\epsilon_z$ (see \Fref{fig:SEM_cant}(b)). For simplicity, we assume that Young's modulus is the same for both layers. We will use this model in all FEM simulations of the various nanomechanical resonators that we analyze (details see \aref{app:FEM}). Further, we measure the relevant properties of the fabricated nanomechancial resonators in a high vacuum chamber ($7 \times 10^{-6}$\,\SI{}{\milli\bar}, if not stated otherwise) at room temperature using an optical interferometric position measurement setup \cite{ciers2024nanomechanical}.

%%%%%%%%%%%%%%%%%%%%%%%%%%%%%%%%%%%%%%%%%%%%%
\subsection{Young's modulus}
%%%%%%%%%%%%%%%%%%%%%%%%%%%%%%%%%%%%%%%%%%%%%
The mechanical frequencies of the modes of a cantilever are given as \cite{schmid2016fundamentals}
\begin{equation}
    f_\text{c} = \frac{\lambda_n^2}{2\pi L^2}\sqrt{\frac{E h^2}{12\rho}},
    \label{eq:f_c}
\end{equation}
where $\lambda_1 = 1.8751$ for the fundamental mode. This formula assumes that the AlN film is vertically uniform, i.e., that Young's modulus and the film's density ($\rho = \SI{3255}{\kilo\gram/\meter}^3$) is the same over the thickness of the cantilever. We will make this simplification here to extract an \textit{effective} Young's modulus, $E$, for our AlN film.
We determine that $E$ is around \SI{310}{\giga\pascal} for AlN films with thicknesses of \SI{90}{\nano\meter}, \SI{145}{\nano\meter}, and \SI{195}{\nano\meter} (\Fref{fig:SEM_cant}(c) and \Tref{tab:h_cant}), but only $\SI{264}{\giga\pascal}$ for the \SI{295}{\nano\meter}-thick AlN film. Note that the latter value is in agreement with Ref.~\cite{ciers2024nanomechanical}, which determined $E = 265 \pm \SI{5}{\giga\pascal}$ from measurements of higher-order modes of uniform beams. We suggest that the lower value of $E$ that is only observed for the \SI{295}{\nano\meter}-thick film could be the result of approaching the critical thickness of the AlN film in our high-temperature MOVPE growth.

%%%%%%%%%%%%%%%%%%%%%%%%%%%%%%%%%%%%%%%%%%%%%
\subsection{Bilayer model}
\label{sec:bilayer}
%%%%%%%%%%%%%%%%%%%%%%%%%%%%%%%%%%%%%%%%%%%%%

In a bilayer structure of layer thicknesses $h_\text{t}$ and $h_\text{b}$, a strain difference $\epsilon_z$ between the layers creates a torque upon release. This torque results in bending of a cantilever with a radius of curvature $R$ of \cite{timoshenko1925analysis}
\begin{equation}
    \frac{1}{R} = \frac{6 \epsilon_z (1+m)^2}{h\left[ 3(1+m)^2 + (1+m) \left( m^2 + 1/m \right) \right]} = \frac{6\epsilon_z h_\text{t} h_\text{b}}{h^3},
    \label{eq:bend}
\end{equation}
where $m = h_\text{t}/h_\text{b}$.

Using a confocal microscope, we measure the bending curvature of \SI{46}{\micro\meter}-, \SI{71}{\micro\meter}-, and \SI{96}{\micro\meter}-long cantilevers near the clamping points.
We fit the data to a circle equation and observe that the bending radius is consistent for cantilevers of a given thickness, see \Fref{fig:SEM_cant}(d). 
This result is expected from \Eref{eq:bend}, as the radius of curvature depends solely on the strain difference and layer thicknesses, but not on the cantilever's length.

\Fref{fig:SEM_cant}(e) shows the bending of AlN cantilevers of various thickness. Overall, the bending radius increases with AlN film thickness (summarized in \Tref{tab:h_cant}). This is because with increasing thickness the contribution of the single-crystal layer to the entire thickness of the AlN film becomes larger, while the thickness of the defect-rich layer remains constant, thus, reducing the overall bending.

\begin{table*}[h!]
    \centering
    \begin{tabular}{c||c c||c|c|c c|c c}
        \hline
        $h$\,(nm) & \multicolumn{2}{c||}{$\sigma$\,(GPa)} & \multicolumn{1}{c|}{$E$\,(GPa)} & $R$\,(\SI{}{\micro\meter}) & \multicolumn{2}{c|}{$f_\text{c}$\,(kHz)} & \multicolumn{2}{c}{$f_\text{b}$\,(MHz)} \\
         & measured & $\sigma_\text{avg}$ & measured & FEM & measured & FEM \\
        \hline
        90 & 0.79 & 0.75 & 313 & 60.8 $\pm$ 0.5 &  67 & 71 &  0.97 & 1.02\\
        145 & 0.88 & 1 & 308 & 67.9 $\pm$ 1.7 & 107 & 112 & 1.1 & 1.18\\
        195 & 1.23 & 1.1 & 328 & 66.4 $\pm$ 0.8 & 149 & 158 & 1.3 & 1.25 \\
        295 &  1.4 & 1.2 &  264 & 100 $\pm$ 1.5& 202 & 218 & 1.4 & 1.3\\
        \hline
    \end{tabular}
    \caption{Residual stress and Young's modulus. The measured residual stress as determined from Raman measurements compared to the average residual stress of the bilayer model used in FEM simulations. The effective Young's modulus is determined from the fundamental mode frequency of cantilevers using \Eref{eq:f_c}.
    $f_\text{c}^\text{meas}$ ($f_\text{c}^\text{FEM}$) and $f_\text{b}^\text{meas}$ ($f_\text{b}^\text{FEM}$) are the measured (FEM simulated) fundamental frequency of \SI{46}{\micro\meter}-long cantilevers and \SI{200}{\micro\meter}-long beams, respectively.}
    \label{tab:h_cant}
\end{table*}

To replicate the bending in the FEM simulations, we model the bottom layer with $h_\text{b}$ = \SI{72.5}{\nano\meter} and a residual stress of $\sigma_{\text{residual,b}}  = \SI{0.6}{\giga\pascal}$, and the top layer with $h_t = h - h_b$ and residual stress of $\sigma_{\text{residual,t}} = \SI{1.4}{\giga\pascal}$. 
The average residual stress, $\sigma_\text{avg}$, is then given as the weighted mean of the two layers, i.e., $\sigma_\text{avg} = (h_\text{b} \sigma_{\text{residual,b}} + h_\text{t} \sigma_{\text{residual,t}})/h$. This value is comparable to the residual stress determined from Raman measurements (see \Tref{tab:h_cant}). \Fref{fig:SEM_cant}(e) compares the cantilever bending obtained from FEM simulations to the measurements. In general, the overall bending behaviour is quite well captured by the FEM simulations. However, the \SI{145}{\nano\meter}- and \SI{195}{\nano\meter}-thick cantilevers bend similarly in the experiment, which is not captured in FEM. This slight discrepancy points to a limitation of the simple bilayer model, which neglects the gradual change in crystal structure that occurs during epitaxial growth.

%%%%%%%%%%%%%%%%%%%%%%%%%%%%%%%%%%%%%%%%%%%%%
\section{Beams}
%%%%%%%%%%%%%%%%%%%%%%%%%%%%%%%%%%%%%%%%%%%%%

\begin{figure*}[t!hbp]
    \centering
    \includegraphics[width=\textwidth]{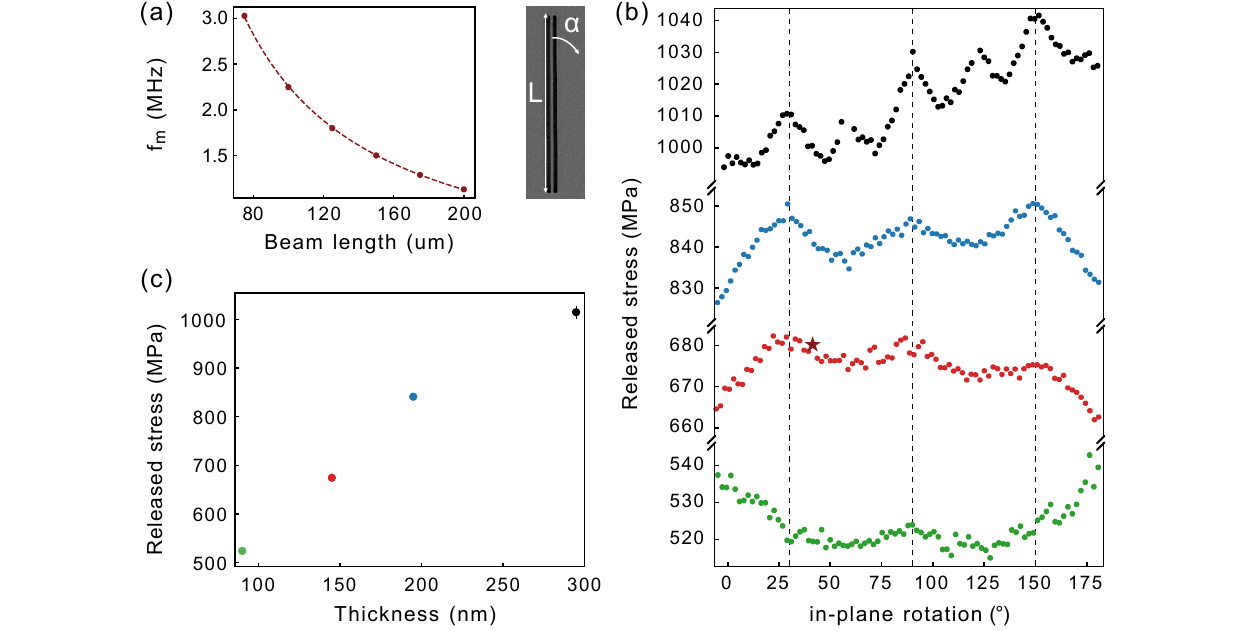}
    \caption{Released stress. (a) Fundamental mode frequency in dependence of beam length in \SI{145}{\nano\meter}-thick AlN. The fit yields \sr{} (star in panel (b)). Right: SEM image of a beam. (b) The dependence of \sr{} on the beam's in-plane orientation. Dashed vertical lines indicate a \SI{60}{\degree} periodicity. (c) Averaged \sr{}.}
    \label{fig:fm_angle}
\end{figure*}

The released stress (\sr{}) and intrinsic quality factor (\Qi{}) of the AlN thin films can be determined from measurements of the eigenfrequencies of uniform beams \cite{unterreithmeier2010damping}. To this end, we have fabricated beams of various lengths (\SI{75}{\micro \meter} to \SI{200}{\micro \meter}) and in-plane orientation angles $\alpha$ (\SI{0}{\degree} to \SI{180}{\degree}), see inset \Fref{fig:fm_angle}(a), as either quantity could be dependent on the orientation of the beam with respect to the crystallographic axes. This is different compared to nanomechanical resonators fabricated from isotropic, e.g., amorphous or polycrystalline, materials. \aref{app:fab} shows images of fabricated beams.

%%%%%%%%%%%%%%%%%%%%%%%%%%%%%%%%%%%%%%%%%%%%%
\subsection{Released stress} 
%%%%%%%%%%%%%%%%%%%%%%%%%%%%%%%%%%%%%%%%%%%%%

For a beam under tensile stress, \sr{}, the fundamental mode frequency is given by \cite{klass2022determining}
\begin{equation}
    f_\text{b} = \frac{1}{2L}\sqrt{\frac{\text{\sr}}{\rho} },
    \label{eq:freq_beam}
\end{equation}
where $L$ is the length of the beam. By measuring the fundamental mode's eigenfrequency of beams of various lengths (\Fref{fig:fm_angle}(a)), we determine the released stress through a fit of the data to \Eref{eq:freq_beam}. \Fref{fig:fm_angle}(b) shows the dependence of the released stress on the orientation of the beam with respect to the crystal axes. We observe a \SI{60}{\degree} periodicity, as is expected from the wurtzite crystal structure of AlN \cite{ciers2024nanomechanical}. For the \SI{295}{\nano\meter}-thick AlN, the released stress is about \SI{1}{\giga\pascal}, whereas for the \SI{90}{\nano\meter}-thick film, the stress decreases to \SI{520}{\mega\pascal}, see \Fref{fig:fm_angle}(c) and \Tref{tab:h_var}. The measured released stress aligns well with the one calculated from the residual stress when accounting for its relaxation via the factor $(1-\nu)$ with the Poisson ratio $\nu = 0.28$.

%%%%%%%%%%%%%%%%%%%%%%%%%%%%%%%%%%%%%%%%%%%%%
\subsection{Intrinsic quality factor}
%%%%%%%%%%%%%%%%%%%%%%%%%%%%%%%%%%%%%%%%%%%%%

The quality factor of a strained, high-aspect-ratio mechanical resonator is enhanced by the dilution factor, $D_Q$, over the intrinsic quality factor, \Qi{} as \cite{fedorov2019generalized}
\begin{equation}
    Q_\text{D} = D_Q \ Q_\text{int}.
    \label{eq:dissipation_dilution}
\end{equation}
While \Qi{} is an intrinsic material property, inversely proportional to the delay between stress and strain, $D_Q$ is determined by the mechanical mode's geometry and can be engineered. The dilution factor depends on the ratio of linear to nonlinear dynamic contributions to the elastic energy, which is specific to the mechanical mode shape \cite{fedorov2019generalized,ciers2024nanomechanical}.

\begin{figure*}[t!hbp]
    \centering
    \includegraphics[width=\textwidth]{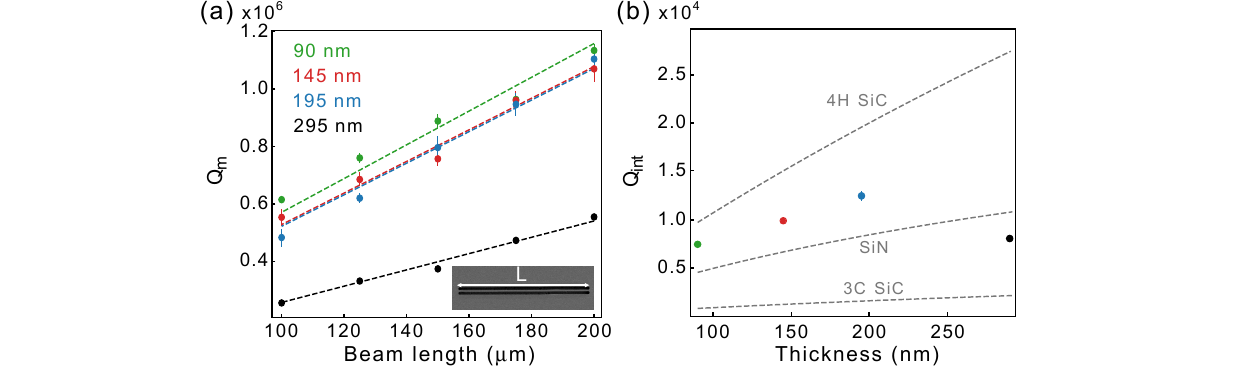}
    \caption{Intrinsic mechanical quality factor of AlN thin films. (a) Data and fit of \Qm{} for \SI{90}{\nano\meter}-, \SI{145}{\nano\meter}-, \SI{195}{\nano\meter}- and \SI{295}{\nano\meter}-thick AlN beams. The inset shows an SEM image of a \SI{200}{\micro\meter}-long beam.
    (b) Calculated \Qi{} using the \er{} values from \Tref{tab:h_var}. The dashed lines show the model of \Qi{} for SiN \cite{villanueva2014evidence}, 3C SiC \cite{romero2020engineering} and 4H SiC \cite{sementilli2024ultralow}.}
    \label{fig:Q_rot}
\end{figure*}

We determine \Qi{} by measuring the quality factor $Q_\text{beam}$ (see \Fref{fig:Q_rot}(a)) of long, thin, strained uniform beams. Their quality factor is limited by dissipation dilution with  $D_{Q}^{\mathrm{beam}}=\left[{(\pi\lambda)^2 + 2\lambda}\right]^{-1}$ and the stress parameter $\lambda = \frac{h}{L} (12\text{\er})^{-1/2}$ \cite{schmid2011damping}. We use the linear relation \er{} = \sr$/E$ (see \Tref{tab:h_var} for the values of $E$) and the values of $Q_\text{beam}$ to determine the intrinsic quality factor. Note that in general the relation between strain and stress is given by a tensor, for details see  Ref.~\cite{ciers2024nanomechanical}. As we do not have experimental access to the elastic constants for each AlN film thickness, but to an effective Young's modulus (see \Eref{eq:f_c}), we make the simplifying assumption of a linear relation between stress and strain. The results are shown in \Fref{fig:Q_rot}(b). We observe that \Qi{} increases with AlN film thickness up to \SI{195}{\nano\meter}-thick films with a value of \Qi{}$=1.24\cdot 10^4$. The thickest AlN film of \SI{295}{\nano\meter}, however, shows a drastic decrease in \Qi{}. A possible reason could be that this film thickness is close to the critical thickness, which results in cracks at the edge of the wafer and, potentially, additional defects in the film, which would lower \Qi{}.

To understand the approximate linear thickness dependence of \Qi{} for AlN films thinner than \SI{200}{\nano\meter}, we compare our results to a commonly-used model for \Qi{}. This model assumes that \Qi{} is determined by bulk loss, $Q_\text{vol}^{-1}$, and surface loss, $Q_\text{s}^{-1}$, as $Q_\text{int}^{-1} = Q_\text{s}^{-1} + Q_\text{vol}^{-1}$  \cite{villanueva2014evidence}. For mechanical resonators with a large surface-to-volume ratio the intrinsic quality factor is predominantly limited by \Qs{}. Surface loss can originate from defects or impurities \cite{villanueva2014evidence}, dangling bonds \cite{mohanty2002intrinsic}, surface roughness \cite{huang2005vhf}, or surface oxidation \cite{tao2015permanent,luhmann2017effect}. 
For example, for SiN nanomechanical resonators Ref.~\cite{villanueva2014evidence} determined  $Q_\text{vol}^{\text{SiN}} = 2.8 \times 10^4$ and ${Q_\text{s}^{\text{SiN}} = \beta_{\text{SiN}}\cdot h}$ with $\beta_{\text{SiN}} = 6 \times 10^{10}$\,1/m, i.e., $Q_\text{s}$ increases linearly with thickness (see \Fref{fig:Q_rot}(b)). We observe a similar behavior for AlN film thicknesses below \SI{200}{\nano\meter} with a slope of $4.8\times 10^{10}$\,1/m. However, unlike SiN, the crystal properties of the AlN thin film change with thickness. We observe, for example, that \Qi{} of the \SI{295}{\nano\meter}-thick AlN film is smaller than the one of the \SI{195}{\nano\meter}-thick film. Thus, we cannot explain that abrupt change by a threshold behavior, which would be given by $Q_\text{vol}$. Overall, the \Qi{} that we determined for AlN thin films up to a film thickness of \SI{200}{\nano\meter} is larger than the corresponding \Qi{} of SiN or epitaxially-grown 3C SiC \cite{romero2020engineering}, but lower than the one of 4H SiC \cite{sementilli2024ultralow} (\Fref{fig:Q_rot}(b)).

\begin{table*}[t!hbp]
    \centering
    \begin{tabular}{c c c c c c}
    \hline
        $h$\,(nm) & \sg{}\,(GPa) & \sr{}\,(GPa) & $E$\,(GPa) & \er{} & \Qi{} \\
        \hline
        90 & 0.79 & 0.52 & 313 & 0.0017 & $7.4 \times 10^3$ \\
        145 & 0.88 & 0.68 & 308 & 0.0022& $9.8 \times 10^3$ \\
        195 & 1.23 & 0.83 & 328 & 0.0025 & $1.24 \times 10^4$ \\
        295 &  1.43 & 1 & 264 & 0.0035 & $8 \times 10^3$ \\
    \hline
    \end{tabular}
    \caption{Mechanical properties of the MOVPE-grown AlN films on Si (111). Residual stress \sg{}, the stress after release \sr{}, and the strain \er{} = \sr$/E$ with Young's modulus $E$. The intrinsic quality factor \Qi{} quantifies the intrinsic mechanical loss.}
    \label{tab:h_var}
\end{table*}

\begin{figure*}[t!hbp]
    \centering
    \includegraphics[width=\textwidth]{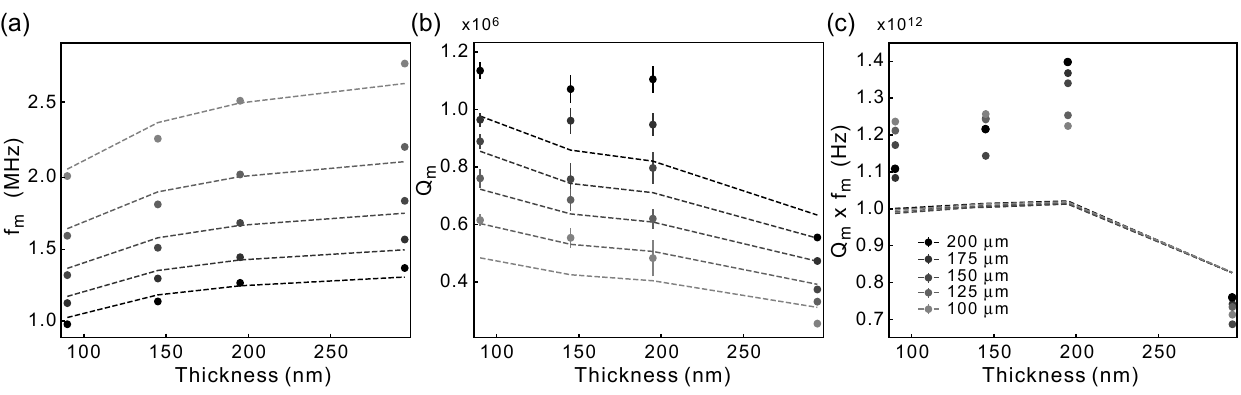}
    \caption{Mechanical eigenfrequency and quality factor of uniform beams fabricated from tensile-strained AlN thin films. (a) Mechanical frequency, \fm{}, (b) mechanical quality factor, \Qm{}, (c) \Qf-product for various thicknesses and beam lengths. Dots are the measured data and dashed lines are the FEM simulation results using the bilayer model. Note that the FEM results for the \Qf-product are independent of the beam's length as expected from the analytical scaling of \fm{} and \Qm{}.}
    \label{fig:beam_Qf}
\end{figure*}

Let us also consider the \Qf-product of the beams, which is a key figure of merit in quantum optomechanics. If \Qf$>6 \times 10^{12}$\,Hz, then the beams would show at least one coherent oscillation at room temperature. The \Qf-product is influenced by the film thickness. While thicker AlN films exhibit higher stress, and, thus, result in a larger mechanical frequency (\Fref{fig:beam_Qf}(a)), their dilution factor decreases (\Fref{fig:beam_Qf}(b)). Further, we observe a linear increase of \Qi{} with thickness only up to \SI{200}{\nano\meter}-thick films. Consequently, there is an optimal film thickness that maximizes the \Qf-product. \Fref{fig:beam_Qf}(c) shows the \Qf-product of beams, where we obtain a maximum value of $1.4\cdot10^{12}\,$Hz at a film thickness of \SI{195}{\nano\meter}, similar to what we expect from FEM simulations using the experimentally determined material properties as input values.

There are several approaches to further improve the intrinsic quality factor, which would directly translate into a larger \Qf-product. For instance, it can be increased by avoiding the defect-rich layer at the interface to the Si substrate as part of the nanomechanical resonator structure.
Therefore, back-etching the \SI{70}{\nano\meter}-thick, defect-rich layer should enhance \Qi{} \cite{romero2020engineering}. 
Another method is wafer-bonding the AlN film to a different substrate and then thinning it down. This approach has been successfully demonstrated with 4H SiC \cite{sementilli2024ultralow}, resulting in a high-quality thin film, but on the cost of reducing the residual stress of the layer. Released beams of wafer-bonded 4H SiC exhibited a stress of \SI{172}{\mega\pascal} only.
Finally, it is possible to introduce a sacrificial layer below the AlN film. This alternation of the growth would relocate the defect-rich interface away from the thin device layer while preserving its stress \cite{kini2020suspended}.

\begin{figure*}[t!hbp]
    \centering
    \includegraphics[width=\textwidth]{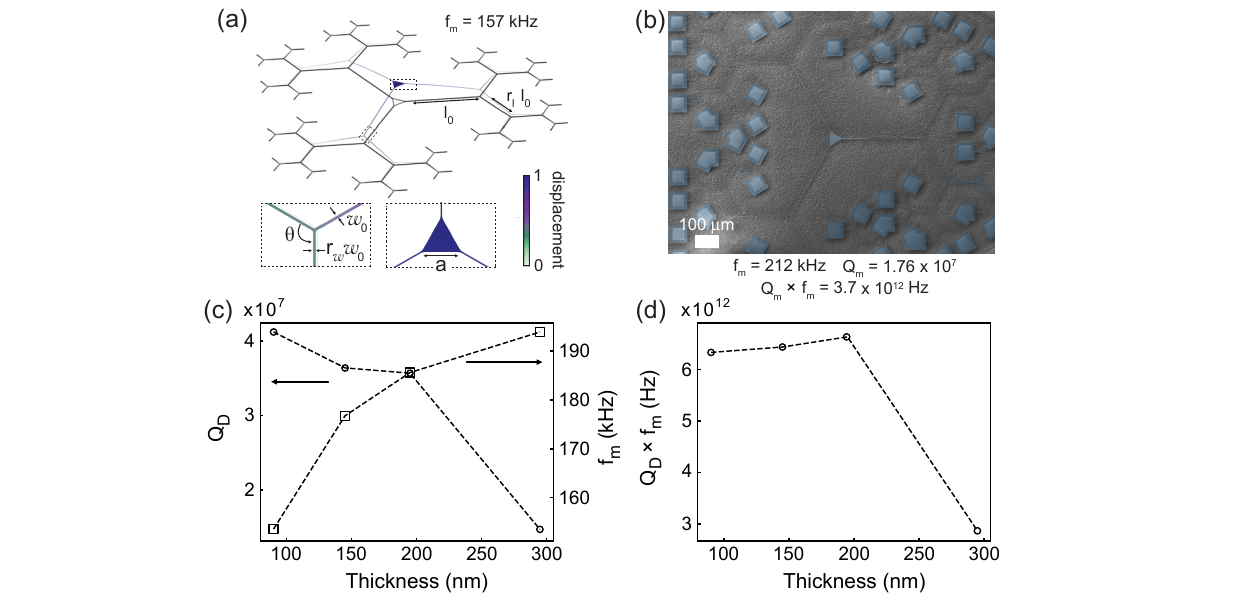}
    \caption{
    Hierarchically-clamped triangline resonators. (a) Schematic of a hierarchically-clamped triangline resonator showing the mechanical displacement of its fundamental mode. (b) False-colored SEM image of a fabricated device. (c) FEM simulated thickness dependence of $Q_{D}$ and \fm{} for the triangline with an equivalent-length of \SI{1.8}{\milli \meter}. (d) FEM simulated values of the $Q_{D} \times f_{m}$-product for the triangline, as a function of thickness. }
    \label{fig:perimeter_triangline}
\end{figure*}

%%%%%%%%%%%%%%%%%%%%%%%%%%%%%%%%%%%%%%%%%%%%%
\section{High-\Qm{} nanomechanical triangline resonators}
%%%%%%%%%%%%%%%%%%%%%%%%%%%%%%%%%%%%%%%%%%%%%

In the following, we demonstrate that the AlN thin films can also be used to realize a complex resonator geometry. To this end, we have chosen a hierarchically-clamped nanomechanical resonator \cite{bereyhi2022hierarchical}, called triangline \cite{ciers2024nanomechanical} (see \Fref{fig:perimeter_triangline}(a)), which exploits dissipation dilution to increase \Qm{}. The triangline has three-fold symmetry and, thus, follows the Al-N bonds of the AlN crystal structure \cite{ciers2024nanomechanical}.
The resonator design features a central pad in the shape of an equilateral triangle with a side length of $a = \SI{70}{\micro \meter}$, suspended by beams of width $w_{0} = \SI{2.0}{\micro \meter}$. At a length of $l_{0} = \SI{0.36}{\milli \meter}$, the string branches in two with an angle of $\theta = \SI{120}{\degree}$ between them. The width of these new branches is given by $w = r_{w} w_{0}$ with $r_{w} =1/(2\cos(\theta/2)) =1$, ensuring that the stress is preserved \cite{bereyhi2022hierarchical}. This branching process repeats $N = 6$ times, with the length of the branches being $l_{0} \cdot  r_{l}^{i}$ for the $i^\textrm{th}$ iteration, where $r_{l} = \SI{0.61}{}$. 
Finally, the last iteration of branches clamps the triangline to the substrate.

We use FEM simulations to predict \Qd{} and \fm{} for the triangline resonator. To this end, we employ the bilayer model with the material parameters that we have experimentally determined. The thickness-dependent values of \Qd{} and \fm{} (\Fref{fig:perimeter_triangline}(b)), and the \Qf{}-product (\Fref{fig:perimeter_triangline}(c)) follow the same trend as was observed for the uniform beams (\Fref{fig:beam_Qf}). The \Qf{}-product of the triangline is expected to lie above $6\cdot 10^{12}$\,Hz when fabricating the triangline in AlN films that are thinner than \SI{200}{\nano\meter}.

We chose to fabricate a triangline in the \SI{90}{\nano \meter}-thick AlN film to demonstrate that the thinnest working AlN material can be used to fabricate such a complex resonator geometry. An SEM image of a fabricated device is shown in \Fref{fig:perimeter_triangline}(b). We observe a fully released resonator structure. We measured a fundamental mode mechanical frequency of $f_\textrm{m} = \SI{212}{\kilo \hertz}$ with a quality factor of $Q_\textrm{m} = \SI{1.76e7}{}$ at a pressure of \SI{4e-8}{\milli \bar}, yielding a \Qf{}-product of $3.7\times 10^{12}$\,Hz, which is slightly lower than the one we expected from FEM simulations. In Ref.~\cite{ciers2024nanomechanical} we realized a triangline of similar length, but in \SI{295}{\nano\meter}-thick AlN and measured a \Qf{}-product of its fundamental mode of $2.5\times 10^{12}$\,Hz. Hence, with the thinner \SI{90}{\nano \meter}-thick AlN film, the \Qf{}-product increased by a factor of 1.5, close to what we expect from FEM simulations (\Fref{fig:perimeter_triangline}(c)).

%%%%%%%%%%%%%%%%%%%%%%%%%%%%%%%%%%%%%%%%%%%%%
\section{Conclusion and outlook}
%%%%%%%%%%%%%%%%%%%%%%%%%%%%%%%%%%%%%%%%%%%%%

We characterized piezoelectric tensile-strained AlN thin films with thicknesses ranging from \SI{45}{\nano\meter} to \SI{295}{\nano\meter} for high-\Qm{} nanomechanics. We identified a defect-rich region of about \SI{70}{\nano\meter} thickness in the AlN film that prevented the release of resonators from films thinner than \SI{90}{\nano\meter}. We determined the highest intrinsic quality factor of $1.2\cdot 10^4$ in the \SI{195}{\nano\meter}-thick AlN film that showed a residual stress of 1.23\,GPa. These values are comparable to other thin film materials used for high-\Qm{} nanomechanics exploiting dissipation dilution. We found that the \Qf-product of AlN films thinner than \SI{200}{\nano\meter} remained nearly constant, as their intrinsic quality factor and stress increased with film thickness, while their dilution factor decreased. Importantly, we showed that the crystalline AlN thin film can be used to realize a complex nanomechanical resonator geometry. To this end, we demonstrated a hierarchically-clamped and triangline-shaped nanomechanical resonator \cite{ciers2024nanomechanical} in the \SI{90}{\nano\meter}-thick AlN film. The fundamental mode of this resonator reached a \Qf-product of $3.7\times 10^{12}$\,Hz, which is 1.5 times larger than the best value achieved with a similar triangline in \SI{295}{\nano\meter}-thick AlN \cite{ciers2024nanomechanical}.

Our results give clear insights on how to increase the \Qf-product in tensile-strained AlN nanomechanics beyond $10^{13}$\,Hz. The MOVPE-based growth of the AlN thin film should be altered such that the defect-rich AlN layer at the interface to the Si substrate is avoided via, e.g., introducing a sacrificial layer \cite{kini2020suspended}. Furthermore, the thickness of the AlN layer shall be well below the critical thickness in MOVPE-based growth \cite{mastro2006mocvd} such that the in-built tensile stress can be maximized. Finally, the growth can be adapted to include doped AlGaN layers \cite{causa2017fernando} around the tensile-strained AlN film such that its piezoelectricity can be harnessed in optoelectromechanical quantum devices \cite{midolo2018nano}.

\begin{acknowledgments}
This work was supported by the Knut and Alice Wallenberg (KAW) Foundation through a Wallenberg Academy Fellowship (W.W.), the KAW project no.~2022.0090, and the Wallenberg Center for Quantum Technology (WACQT, A.C.), and the Swedish Research Council (VR project No. 2019-04946). H.P.~acknowledges funding by the European Union under the project MSCA-PF-2022-OCOMM. MOVPE of AlN on Si was performed at Otto-von-Guericke-University Magdeburg. The mechanical resonators were fabricated in the Myfab Nanofabrication Laboratory at Chalmers and analyzed in Chalmers Materials Analysis Laboratory. 

\end{acknowledgments}

\section*{Data availability}
Data underlying the results presented in this paper are available in the open-access Zenodo database: \href{https://doi.org/10.5281/zenodo.13890719}{10.5281/zenodo.13890719} \cite{zenododata}.

\section*{Disclosures}
Witlef Wieczorek is a shareholder in WACQT-IP AB.

%%%%%%%%%%%%%%%%%%%%%%%%%%%%%%%%%%%%%%%%%%%%%
%\clearpage

\appendix

%%%%%%%%%%%%%%%%%%%%%%%%%%%%%%%%%%%%%%%%%%%%%
%%%%%%%%%%%%%%%%%%%%%%%%%%%%%%%%%%%%%%%%%%%%%

%%%%%%%%%%%%%%%%%%%%%%%%%%%%%%%%%%%%%%%%%%%%%
%%%%%%%%%%%%%%%%%%%%%%%%%%%%%%%%%%%%%%%%%%%%%
\section{Methods}\label{app:app}
%%%%%%%%%%%%%%%%%%%%%%%%%%%%%%%%%%%%%%%%%%%%%
\subsection{Growth}\label{app:growth}
%%%%%%%%%%%%%%%%%%%%%%%%%%%%%%%%%%%%%%%%%%%%%
The AlN film was grown by metal-organic vapour-phase epitaxy (MOVPE, AIXTRON AIX 200/4 RF-S) on a 2-inch \SI{500}{\micro\meter}-thick highly As-doped silicon (111) wafer.
After a thin Al deposition, the growth of AlN was performed in two regimes.
First \SI{20}{\nano\meter} were grown at a low growth rate with a surface temperature of \SI{1110}{\degree}C, \SI{100}{\milli\bar}, and a high V-III ratio of 2500.
Then the main AlN layer was grown at a high growth rate at \SI{70}{\milli\bar} and a low V-III ratio of 25 with a surface temperature of \SI{1120}{\degree}C. By varying the time of the main deposition, samples with different AlN thicknesses were produced.

\subsection{Material characterization}\label{app:char}

\subsubsection{XRD measurements}

\begin{figure*}[t!hbp]
    \centering
    \includegraphics[width=\textwidth]{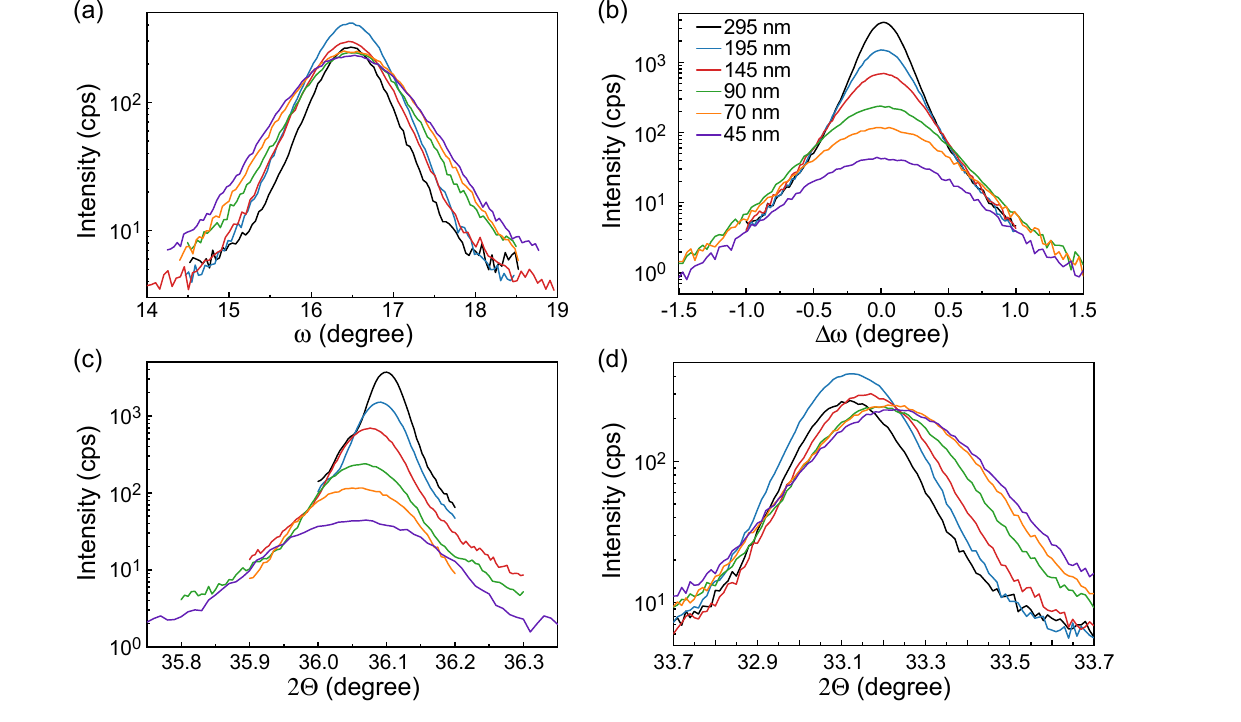}
    \caption{(a) Grazing incidence in-plane diffraction (GIID) $\omega$-scan for ($10\bar{1}0$). (b) $\omega$-scan at ($0002$). (c) $2\Theta/\theta$-scan for ($0002$). (d) GIID $2\Theta/\theta$-scan for ($10\bar{1}0$).}
    \label{fig:XRD_scans}
\end{figure*}

High resolution X-Ray diffraction (HRXRD) measurements were performed to obtain information about the crystallographic structure and orientation of the AlN thin films.
In an $\omega$-scan the incident X-Ray beam's angle to the selected lattice plane is varied, while the detector angle $2\Theta$ is fixed. Such scan can be performed in a grazing incidence in-plane diffraction (GIID) setup. The measurement in such a configuration is shown in \Fref{fig:XRD_scans}(a). Grazing incidence means that the incident X-ray beam comes in at a very shallow angle (close to parallel) to the surface and the measured diffraction is collected in the plane of the sample surface. This scan gives information on in-plane orientation disorder of the crystallites ($c$-axis twist).

Another useful measurement of the $c$-axis disorder (tilt) can be performed by a slight variation of the incident $\omega$ angle when the detector angle is fixed to the symmetric lattice plane (0002). The measurement is shown in \Fref{fig:XRD_scans}(b).
When the ratio between the maximium of the $\omega$-scan ($\omega=\Theta$) and detector is kept constant during the scan ($2\Theta/\theta$-scan), the lattice parameters are determined (i.e., the $c$ value), see the measurement in \Fref{fig:XRD_scans}(c).
To examine the in-plane lattice parameter (i.e., $a$), a $2\Theta/\theta$-scan is performed in GIID setting (on 10$\bar{1}$0), see the measurement in \Fref{fig:XRD_scans}(d).

\subsubsection{Raman measurements}
Raman scattering is measured using a confocal Raman microscope at room temperature with a \SI{532}{\nano\meter} laser for excitation. The laser beam is focused by an objective lens with a magnification of 100$\times$ and a numerical aperture of 0.9, leading to an optical spot size of about \SI{300}{\nano \meter}. We verified that the Raman signal is power independent and, thus, unaffected by a potential heating of the device.

\subsection{Fabrication process}\label{app:fab}

\begin{figure*}[t!hbp]
    \centering
    \includegraphics[width=\textwidth]{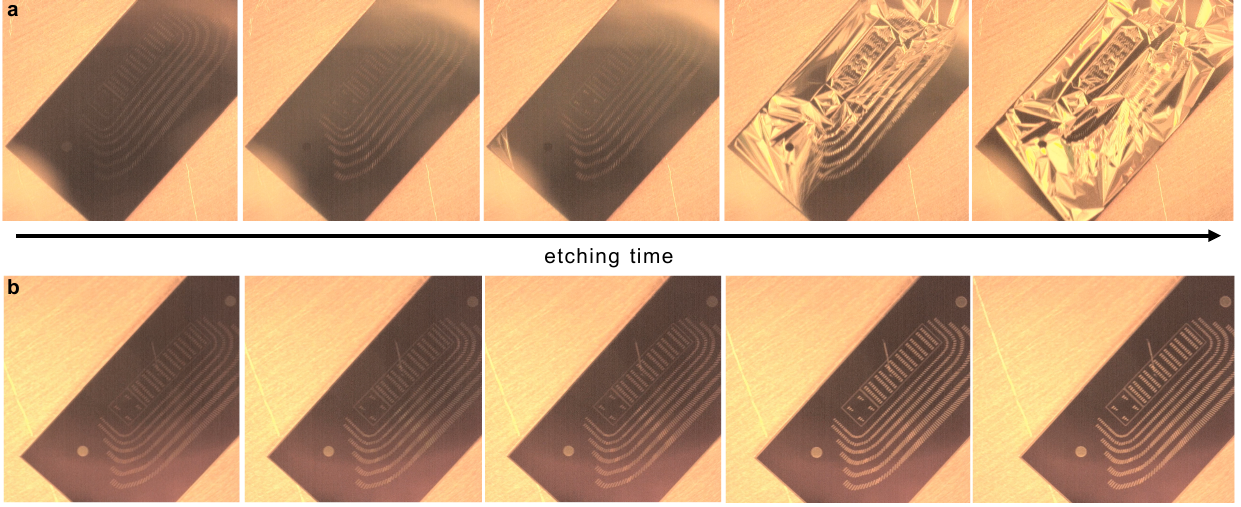}
    \caption{Time-dependent release process of AlN thin films. Temporal snapshots of the XeF$_2$ etching process of $5 \times \SI{10}{\milli\meter}$ chips for (a) \SI{70}{\nano\meter}-thick AlN, and (b) \SI{195}{\nano\meter}-thick AlN.}
    \label{fig:pilling}
\end{figure*}

\begin{figure*}[t!hbp]
    \centering
    \includegraphics[width=\textwidth]{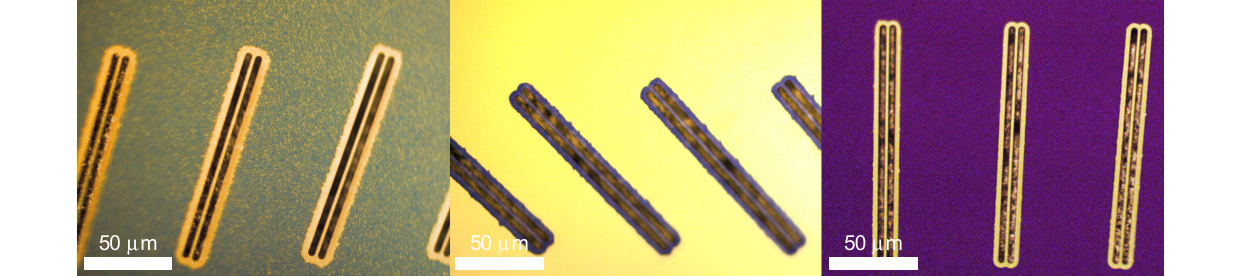}
    \caption{Optical microscope images of uniform beams fabricated from \SI{90}{\nano\meter}-, \SI{145}{\nano\meter}- and \SI{195}{\nano\meter}-thick AlN.}
    \label{fig:opt_beams}
\end{figure*}

The fabrication of nanomechanical resonators begins by sputtering a \SI{50}{\nano\meter} SiO$_2$ hard mask. Subsequently, we define the pattern of the mechanical resonator in electron-beam resist (UV-60). 
Then we transfer the pattern into the hard mask and AlN film in consequent ICP/RIE etching steps with CF$_4$/CHF$_3$ and Cl$_2$/Ar mixtures, respectively. 
We strip the photoresist with NMP (Remover 1165), and the sample is cleaned with one minute HF etching.
When the AlN film is suspended by etching Si with XeF$_2$, \SI{45}{\nano\meter}- and \SI{70}{\nano\meter}-thick AlN peeled off the substrate, see \Fref{fig:pilling}(a).
This indicates that such films mostly consist of a defect-rich layer. Instead, AlN films with a thickness above \SI{90}{\nano\meter} can be successfully suspended, \Fref{fig:pilling}(b). Note that we observed that the \SI{90}{\nano\meter}-thick AlN film is grainy after XeF$_2$ etching (highest RMS roughness of all films, see \Tref{tab:AFM_h}), while thicker films are not (see \Fref{fig:opt_beams}).

\subsection{Finite element simulations}\label{app:FEM}

\begin{figure*}[t!hbp]
    \centering
    \includegraphics[width=\textwidth]{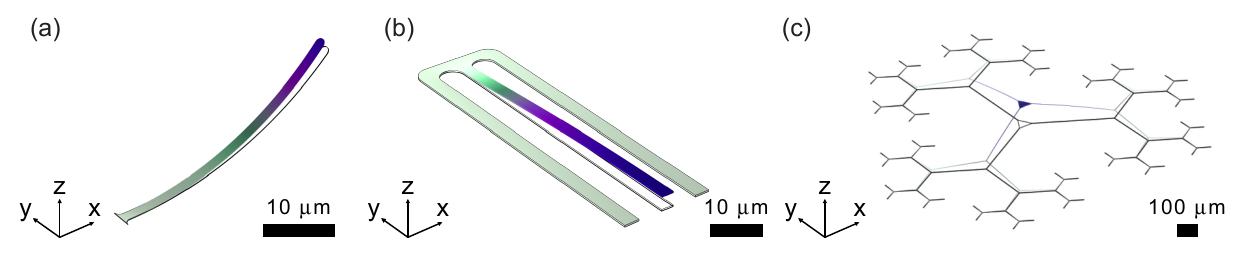}
    \caption{FEM simulations of the mechanical displacement for the fundamental out-of-plane oscillation mode of (a) cantilever, (b) beam (mirror symmetry is applied), and (c) triangline.}
    \label{fig:FEM_3D}
\end{figure*}

We use the solid mechanics interface of COMSOL Multiphysics for FEM simulations and show examples of simulated devices in \Fref{fig:FEM_3D}. Details of the simulations can be found in Ref.~\cite{ciers2024nanomechanical}. Here, we assume an isotropic material in our simulations as we use the experimentally determined values, among them Young's modulus, \Qi{}, and stress (see \Tref{tab:h_var}). Additionally, we use as parameters an AlN density of \SI{3255}{\kilo\gram/\meter}$^3$ and Poisson's ratio of 0.28.

\clearpage
%\nocite{*}
\bibliography{biblio}

\end{document}